\documentclass[prd,reprint,showpacs,showkeys]{revtex4-1}
\usepackage{amsfonts}
\usepackage{amssymb}
\usepackage{amsmath}
\usepackage{graphicx}
\usepackage[font={footnotesize,it}]{caption}

\begin{document}

\title{Microscopic thin shell wormholes in magnetic Melvin universe}
\author{S. Habib Mazharimousavi}
\email{habib.mazhari@emu.edu.tr}
\author{M. Halilsoy}
\email{mustafa.halilsoy@emu.edu.tr}
\author{Z. Amirabi}
\email{zahra.amirabi@emu.edu.tr}
\affiliation{Department of Physics, Eastern Mediterranean University, G. Magusa, north
Cyprus, Mersin 10, Turkey. }
\date{\today }

\begin{abstract}
We construct thin shell wormholes in the magnetic Melvin universe. It is
shown that in order to make a TSW in the Melvin spacetime the radius of the
throat can not be larger than $\frac{2}{B_{0}}$ in which $B_{0}$ is the
magnetic field constant. We also analyze the stability of the constructed
wormhole in terms of a linear perturbation around the equilibrium point. In
our stability analysis we scan a full set of the Equation of States such as
Linear Gas, Chaplygin Gas, Generalized Chaplygin Gas, Modified Generalized
Chaplygin Gas and Logarithmic Gas. Finally we extend our study to the
wormhole solution in the unified Melvin and Bertotti-Robinson spacetime. In
this extension we show that for some specific cases, the local energy
density is partially positive but the total energy which supports the
wormhole is positive.
\end{abstract}

\maketitle

\section{Introduction}

The magnetic Melvin universe (more appropriately the Bonnor-Melvin universe) 
\cite{Melvin} is sourced by a beam of magnetic field parallel to the $z$%
-axis in the Weyl coordinates \{$t,\rho ,z,\varphi $\}. The metric depends
only on the radial coordinate $\rho $ which makes a typical case of
cylindrical symmetry. It is a regular, non-black hole solution of the
Einstein-Maxwell equations. Behaviour of the magnetic field is $B\left( \rho
\right) \sim \rho $ (for $\rho \rightarrow 0$) and $B\left( \rho \right)
\sim \frac{1}{\rho ^{3}}$ (for $\rho \rightarrow \infty $). At radial
infinity the magnetic field vanishes but spacetime is not flat. On the
symmetry axis ($\rho =0$) the magnetic field vanishes; since the behaviour
is same for $0\leqslant \left\vert z\right\vert <\infty $ the Melvin
spacetime is not asymptotically flat also for $\left\vert z\right\vert
\rightarrow \infty $. The magnetic field can be assumed strong enough to
warp spacetime to the extent that it produces possible wormholes. Strong
magnetic fields are available in magnetars (i.e. $B\sim 10^{15}G,$ while our
Earth's magnetic field is $B_{Earth}\sim 0.5G$), pulsars and other objects.
Since creation of strong magnetic fields can be at our disposal in a
laboratory - at least in very short time intervals - it is natural to raise
the question whether wormholes can be produced in a magnetized
superconducting environment. \textit{From this reasoning we aim to construct
a thin-shell wormhole (TSW) in a magnetic Melvin universe.} The method is an
art of spacetime tailoring, i.e. cutting and pasting at a throat region
under well-defined mathematical junction conditions. Some related papers can
be found in \cite{TSWS} for spherically symmetric bulk and in \cite{TSWC}
for cylindrically symmetric. The TSW\ is threaded by exotic matter which is
taken for granted, and our principal aim is to search for the stability
criteria for such a wormhole. Two cylindrically symmetric Melvin universes
are glued at a hypersurface radius $\rho =a=$constant, which is endowed with
surface energy-momentum to provide necessary support against the
gravitational collapse. It turns out that in the Melvin spacetime the radial
flare-out condition, i.e. $\frac{dg_{\varphi \varphi }}{da}>0$ is satisfied
for a restricted radial distance, which makes a small scale wormhole.
Specifically, this amounts to a throat radius $\rho =a<\frac{2}{\left\vert
B_{o}\right\vert },$ so that for high magnetic fields the throat radius can
be made arbitrarily small. This can be dubbed as a microscopic wormhole. As
stated recently such small wormholes may host the quantum
Einstein-Podolsky-Rosen (EPR) pair \cite{EPR}. The throat is linearly
perturbed in the radial distance and the resulting perturbation equation is
obtained. The problem is reduced to a one-dimensional particle problem whose
oscillatory behavior for an effective potential $V(a)$ about the equilibrium
point is provided by $V^{\prime \prime }(a_{0})>0$. Given the Equation of
State (EoS) on the hypersurface we plot the parametric stability condition $%
V^{\prime \prime }(a_{0})>0$ to determine the possible stable regions. Our
samples of EoS consist of a Linear Gas, various forms of Chaplygin Gas and a
Logarithmic gas. We consider TSW also in the recently found
Melvin-Bertotti-Robinson magnetic universe \cite{MH1}. In the
Bertotti-Robinson limit the wormhole is supported by total positive energy
for any finite extension in the axial direction. For infinite extension the
total energy reduces to zero, at least better than the total negative
classical energy.

Organization of the paper is as follows. Construction of TSW from the
magnetic Melvin spacetime is introduced in Sec. II. Stability of the TSW is
discussed in Sec. III. Sec. IV discusses the consequences of small velocity
perturbations. Section V considers TSW in Melvin-Bertotti-Robinson spacetime
and Conclusion in Sec. VI completes the paper.

\section{Thin-shell wormhole in Melvin geometry}

Let's start with the Melvin magnetic universe spacetime \cite{Melvin} in its
axially symmetric form%
\begin{equation}
ds^{2}=U\left( \rho \right) \left( -dt^{2}+d\rho ^{2}+dz^{2}\right) +\frac{%
\rho ^{2}}{U\left( \rho \right) }d\varphi ^{2}
\end{equation}%
in which 
\begin{equation}
U\left( \rho \right) =\left( 1+\frac{B_{0}^{2}}{4}\rho ^{2}\right) ^{2}
\end{equation}%
where $B_{0}$ denotes the magnetic field constant. The Maxwell field
two-form, however, is given by%
\begin{equation}
\mathbf{F}=\frac{\rho B_{0}}{U\left( \rho \right) }d\rho \wedge d\varphi .
\end{equation}%
We note that the Melvin solution in Einstein-Maxwell theory does not
represent a black hole solution. The solution is regular everywhere as seen
from the Ricci scalar and Ricci sequence 
\begin{eqnarray}
R &=&0 \\
R_{\mu \nu }R^{\mu \nu } &=&\frac{4B_{0}^{4}}{U\left( \rho \right) ^{8}} 
\notag
\end{eqnarray}%
as well as the Kretschmann scalar%
\begin{equation}
\mathcal{K}=\frac{4B_{0}^{4}\left( 3B_{0}^{4}\rho ^{4}-24B_{0}^{2}\rho
^{2}+80\right) }{U\left( \rho \right) ^{8}}.
\end{equation}%
In \cite{BR}, the general conditions which should be satisfied to have
cylindrical wormhole possible are discussed. In brief, while the stronger
condition implies that $\sqrt{g_{\varphi \varphi }}$ should take its minimum
value at the throat, the weaker condition states that $\sqrt{g_{\varphi
\varphi }g_{zz}}$ should be minimum at the throat. The stronger and weaker
conditions are called \textit{radial flare-out and areal flare-out
conditions }respectively \cite{ES1,ES2,R}. As we shall see in the sequel, in
the case of TSW $\sqrt{g_{\varphi \varphi }}$ and $\sqrt{g_{\varphi \varphi
}g_{zz}}$ should only be increasing function at the throat in \textit{radial
flare-out and areal flare-out conditions.} In the case of the Melvin
spacetime, 
\begin{equation}
\sqrt{g_{\varphi \varphi }}=\frac{\rho }{1+\frac{B_{0}^{2}}{4}\rho ^{2}}
\end{equation}%
and 
\begin{equation}
\sqrt{g_{\varphi \varphi }g_{zz}}=\rho .
\end{equation}%
One easily finds that \textit{areal flare-out condition }is trivially
satisfied and the \textit{radial flare-out condition} requires $\rho <\frac{2%
}{B_{0}}$. \textit{\ }

Following Visser \cite{Visser}, from the bulk spacetime (1) we cut two
non-asymptotically flat copies $\mathcal{M}^{\pm }$ from a radius $\rho =a$
with $a>0$ and then we glue them at a hypersurface $\Sigma =\Sigma ^{\pm }$
which is defined as $\mathcal{H}\left( \rho \right) =\rho -a\left( \tau
\right) =0.$ In this way the resultant manifold is complete. At hypersurface 
$\Sigma $ the induced line element is given by%
\begin{equation}
ds^{2}=-d\tau ^{2}+U\left( a\right) dz^{2}+\frac{a^{2}}{U\left( a\right) }%
d\varphi ^{2}
\end{equation}%
in which 
\begin{equation}
-1=U\left( a\right) \left( -\dot{t}^{2}+\dot{\rho}^{2}\right)
\end{equation}%
where a dot stands for derivative with respect to the proper time $\tau $ on
the hypersurface $\Sigma $. The Israel junction conditions which are the
Einstein equations on the junction hypersurface read as ($8\pi G=1$)%
\begin{equation}
k_{i}^{j}-k\delta _{i}^{j}=-S_{i}^{j},
\end{equation}%
in which $k_{i}^{j}=K_{i}^{j\left( +\right) }-K_{i}^{j\left( -\right) },$ $%
k=tr\left( k_{i}^{j}\right) $ and 
\begin{equation}
K_{ij}^{\left( \pm \right) }=-n_{\gamma }^{\left( \pm \right) }\left( \frac{%
\partial ^{2}x^{\gamma }}{\partial X^{i}\partial X^{j}}+\Gamma _{\alpha
\beta }^{\gamma }\frac{\partial x^{\alpha }}{\partial X^{i}}\frac{\partial
x^{\beta }}{\partial X^{j}}\right) _{\Sigma }
\end{equation}%
is the extrinsic curvature. Also the normal unit vector is defined as 
\begin{equation}
n_{\gamma }^{\left( \pm \right) }=\left( \pm \left\vert g^{\alpha \beta }%
\frac{\partial \mathcal{H}}{\partial x^{\alpha }}\frac{\partial \mathcal{H}}{%
\partial x^{\beta }}\right\vert ^{-1/2}\frac{\partial \mathcal{H}}{\partial
x^{\gamma }}\right) _{\Sigma }
\end{equation}%
and $S_{i}^{j}=$diag$\left( -\sigma ,P_{z},P_{\varphi }\right) $ is the
energy momentum tensor on $\Sigma $. Explicitly we find, 
\begin{equation}
n_{\gamma }^{\left( \pm \right) }=\pm \left( -\dot{a}U\left( a\right)
,U\left( a\right) \sqrt{\Delta },0,0\right) _{\Sigma },
\end{equation}%
in which $\Delta =\frac{1}{U\left( a\right) }+\dot{a}^{2}$. The non-zero
components of the extrinsic curvature are found as 
\begin{equation}
K_{\tau }^{\tau \left( \pm \right) }=\pm \frac{1}{\sqrt{\Delta }}\left( 
\ddot{a}+\frac{U^{\prime }}{U}\dot{a}^{2}+\frac{U^{\prime }}{2U^{2}}\right)
\end{equation}%
\begin{equation}
K_{z}^{z\left( \pm \right) }=\pm \frac{U^{\prime }}{2U}\sqrt{\Delta },
\end{equation}%
and%
\begin{equation}
K_{\varphi }^{\varphi \left( \pm \right) }=\pm \left( \frac{1}{a}-\frac{%
U^{\prime }}{2U}\right) \sqrt{\Delta },
\end{equation}%
in which prime implies $\frac{\partial }{\partial a}$. Imposing the junction
conditions \cite{Israel} we find the components of the energy momentum
tensor on the shell which are expressed as 
\begin{equation}
\sigma =-\frac{2}{a}\sqrt{\Delta }
\end{equation}%
\begin{equation}
P_{z}=\frac{2\ddot{a}+\frac{2U^{\prime }}{U}\dot{a}^{2}+\frac{U^{\prime }}{%
U^{2}}}{\sqrt{\Delta }}+\left( \frac{2}{a}-\frac{U^{\prime }}{U}\right) 
\sqrt{\Delta },
\end{equation}%
and%
\begin{equation}
P_{\varphi }=\frac{2\ddot{a}+\frac{2U^{\prime }}{U}\dot{a}^{2}+\frac{%
U^{\prime }}{U^{2}}}{\sqrt{\Delta }}+\frac{U^{\prime }}{U}\sqrt{\Delta }.
\end{equation}%
Having energy density on the shell, one may find the total exotic matter
which supports the wormhole per unit $z$ by%
\begin{equation}
\Omega =2\pi aU\left( a\right) \sigma
\end{equation}%
which is clearly exotic.

\section{Stability of the thin-shell wormhole against a linear perturbation}

Recently, we have generalized the stability of TSWs in cylindrical symmetric
bulks in \cite{MH2}. Here we apply the same method to the TSWs in Melvin
universe. Similar to the spherical symmetric TSW, we start with the energy
conservation identity on the shell which implies%
\begin{multline}
\left( aS_{;j}^{ij}=\right) \frac{d}{d\tau }\left( a\sigma \right) +\left[ 
\frac{aU^{\prime }}{2U}\left( P_{z}-P_{\varphi }\right) +P_{\varphi }\right] 
\frac{da}{d\tau } \\
=\frac{da}{d\tau }\frac{U^{\prime }}{U}\left( 4-a\frac{U^{\prime }}{U}%
\right) \sqrt{\Delta }.
\end{multline}

As we have shown in previous section the expressions given for surface
energy density $\sigma $ and surface pressures $P_{z}$ and $P_{\varphi }$
are for a dynamic wormhole. This means that if there exists an equilibrium
radius for the throat radius, say $a=a_{0},$ at this point $\dot{a}_{0}=0$
and $\ddot{a}_{0}=0$ and consequently the form of the surface energy density
and pressure reduce to the static forms as%
\begin{equation}
\sigma _{0}=-\frac{2}{a_{0}\sqrt{U_{0}}}
\end{equation}%
\begin{equation}
P_{z0}=\frac{2}{a_{0}\sqrt{U_{0}}}
\end{equation}%
and%
\begin{equation}
P_{\varphi 0}=2\frac{U_{0}^{\prime }}{U_{0}\sqrt{U_{0}}}.
\end{equation}%
Let's assume that after the perturbation the surface pressures are a general
function of $\sigma $ which may be written as 
\begin{equation}
P_{z}=\Psi \left( \sigma \right)
\end{equation}%
and%
\begin{equation}
P_{\varphi }=\Phi \left( \sigma \right)
\end{equation}%
such that at the throat i.e. $a=a_{0}$, $\Psi \left( \sigma _{0}\right) =$ $%
P_{z0}$ and $\Phi \left( \sigma _{0}\right) =P_{\varphi 0}.$ From (17) one
finds a one-dimensional type equation of motion for the throat 
\begin{equation}
\dot{a}^{2}+V\left( a\right) =0
\end{equation}%
in which $V\left( a\right) $ is given by 
\begin{equation}
V\left( a\right) =\frac{1}{U}-\left( \frac{a\sigma }{2}\right) ^{2}.
\end{equation}%
Using the energy conservation identity (21), one finds%
\begin{multline}
\left( a\sigma \right) ^{\prime }=-\left[ \frac{aU^{\prime }}{2U}\left( \Psi
\left( \sigma \right) -\Phi \left( \sigma \right) \right) +\Phi \left(
\sigma \right) \right] + \\
\frac{U^{\prime }}{U}\left( 4-a\frac{U^{\prime }}{U}\right) \sqrt{\Delta },
\end{multline}%
which helps us to show that $V^{\prime }\left( a_{0}\right) =0$ and

\begin{multline}
V_{0}^{\prime \prime }=\frac{\left( 2U_{0}+a_{0}U_{0}^{\prime }\right) \left[
U_{0}^{\prime }\left( \Phi _{0}^{\prime }-\Psi _{0}^{\prime }\right)
a_{0}-2U_{0}\Phi _{0}^{\prime }\right] }{2U_{0}^{3}a_{0}^{2}}- \\
\frac{U_{0}^{2}\left( 2U_{0}^{\prime }-4a_{0}U_{0}^{\prime \prime }\right)
+U_{0}\left( 2U_{0}^{\prime \prime }U_{0}^{\prime
}a_{0}^{2}+7a_{0}U_{0}^{\prime 2}\right) -3a_{0}^{2}U_{0}^{\prime 3}}{%
2U_{0}^{4}a_{0}}.
\end{multline}%
Note that a sub zero means that the corresponding quantity is evaluated at
the equilibrium radius i.e., $a=a_{0}$. We also note that a prime denotes
derivative with respect to its argument, for instance $\Psi _{0}^{\prime
}=\left. \frac{\partial \Psi }{\partial \sigma }\right\vert _{\sigma =\sigma
_{0}}$ while $U_{0}^{\prime }=\left. \frac{\partial U}{\partial a}%
\right\vert _{a=a_{0}}.$ Now, if we expand the equation of motion of the
throat about $a=a_{0}$ we find (up to second order)%
\begin{equation}
\ddot{x}+\omega ^{2}x\tilde{=}0
\end{equation}%
in which $x=a-a_{0}$ and $\omega ^{2}=\frac{1}{2}V^{\prime \prime }\left(
a_{0}\right) .$ This equation describes the motion of a harmonic oscillator
provided $\omega ^{2}>0$ which is the case of stability. If $\omega ^{2}<0$
it implies that after the perturbation an exponential form fails to return
back to its equilibrium point and therefore the wormhole is called unstable.

To conclude about the stability of the TSW in Melvin magnetic space we
should examine the sign of $V^{\prime \prime }\left( a_{0}\right) $ and in
any region where $V^{\prime \prime }\left( a_{0}\right) >0$ the wormhole is
stable and in contrast if $V^{\prime \prime }\left( a_{0}\right) <0$ we
conclude that the wormhole is unstable. From Eq. (30), we observe that this
issue is identified with $a,$ $U_{0},$ $U_{0}^{\prime },$ $U_{0}^{\prime
\prime }$ together with $\Phi _{0}^{\prime }$ and $\Psi _{0}^{\prime }.$
Since the form of $U\left( a\right) $ is known in order to examine the
stability of the wormhole one should choose a specific EoS i.e. $\Psi \left(
\sigma \right) $ and $\Phi \left( \sigma \right) $. In the following chapter
we shall consider the well known cases of EoS which have been introduced in
the literature. For each case we determine whether the TSW is stable or not.

\subsection{Specific EoS}

As we have already mentioned, in this chapter we go through the details of
some specific EoS and the stability of the corresponding TSW.

\subsubsection{Linear Gas (LG)}

\begin{figure}[tbp]
\includegraphics[width=60mm,scale=0.7]{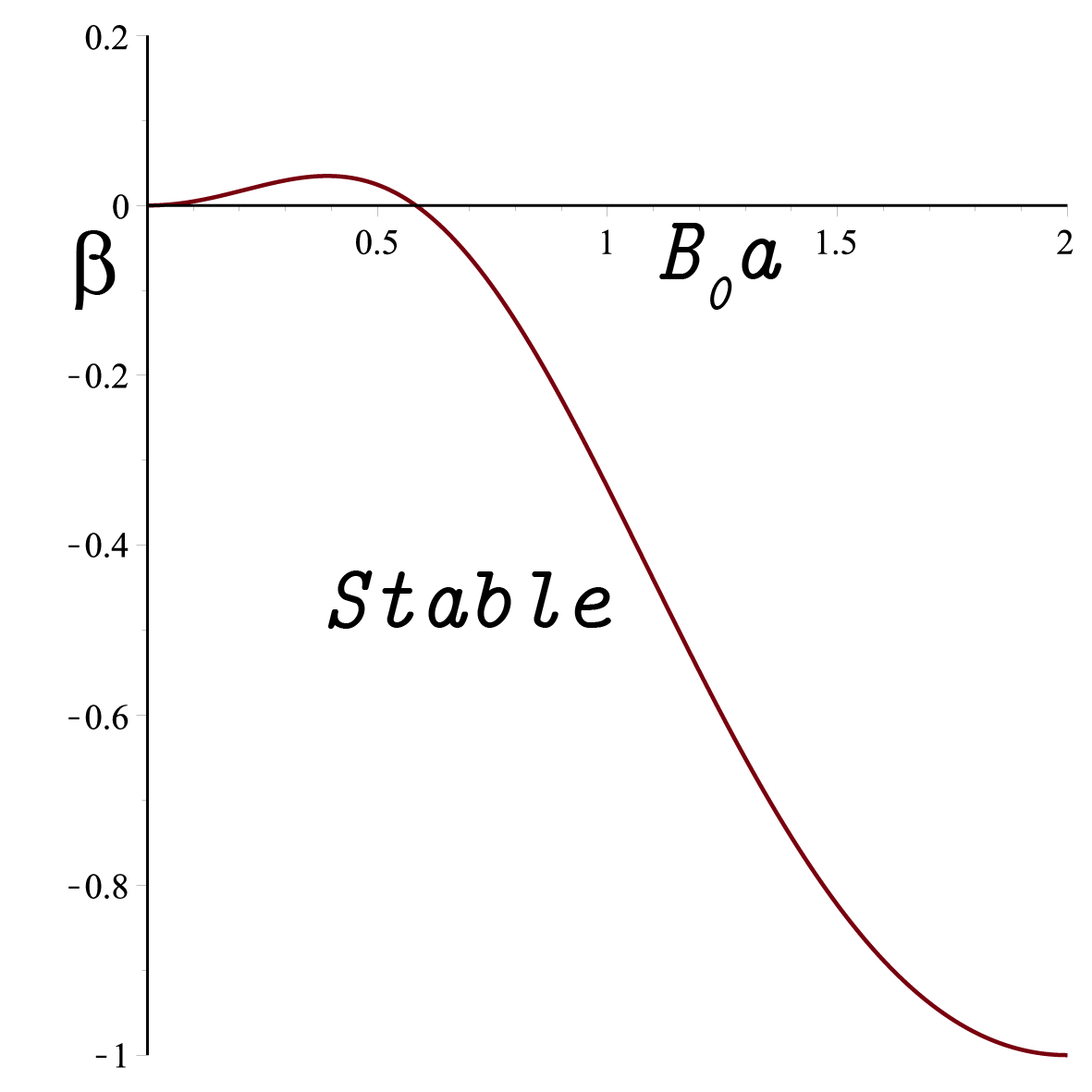}
\caption{Stability of TSW supported by LG in terms of $a_{0}B_{0}$ and $%
\protect\beta =\protect\beta _{1}=\protect\beta _{2}$. We note that the
upper bound of $a_{0}B_{0}$ is chosen to be $2.$ This let $\frac{a^{2}}{%
f\left( a\right) }$ to remain an increasing function with respect to $a.$
This condition is needed to have a TSW possible in CS spacetime [5].}
\end{figure}

Our first choice of the EoS is a LG in which $\Psi ^{\prime }\left( \sigma
\right) =\beta _{1}$ and $\Phi ^{\prime }\left( \sigma \right) =\beta _{2}$
with $\beta _{1}$ and $\beta _{2}$, two constant parameters related to the
speed of sound in $z$ and $\varphi $ directions. We also find the form of $%
\Psi \left( \sigma \right) $ and $\Phi \left( \sigma \right) $ which are%
\begin{equation}
\Psi \left( \sigma \right) =\beta _{1}\sigma +\Psi _{0}
\end{equation}%
and%
\begin{equation}
\Phi \left( \sigma \right) =\beta _{2}\sigma +\Phi _{0}
\end{equation}%
with $\Psi _{0}$ and $\Phi _{0}$ as integration constants. We impose $\Psi
\left( \sigma _{0}\right) =P_{z0}$ and $\Phi \left( \sigma _{0}\right)
=P_{\varphi 0},$ which yields%
\begin{equation}
\Psi _{0}=P_{z0}-\beta _{1}\sigma _{0}
\end{equation}%
and%
\begin{equation}
\Phi _{0}=P_{\varphi 0}-\beta _{2}\sigma _{0}.
\end{equation}%
In the case with $\beta _{1}=$ $\beta _{2}=\beta $, we find that $\Psi $ and 
$\Phi $ are related as 
\begin{equation}
\Psi -\Phi =P_{z0}-P_{\varphi 0}
\end{equation}%
but in general they are independent. In Fig. 1 we consider $\beta _{1}=$ $%
\beta _{2}=\beta $ and the resulting stable region with $V_{0}^{\prime
\prime }>0$ is displayed.

\subsubsection{Chaplygin Gas (CG)}

\begin{figure}[tbp]
\includegraphics[width=60mm,scale=0.7]{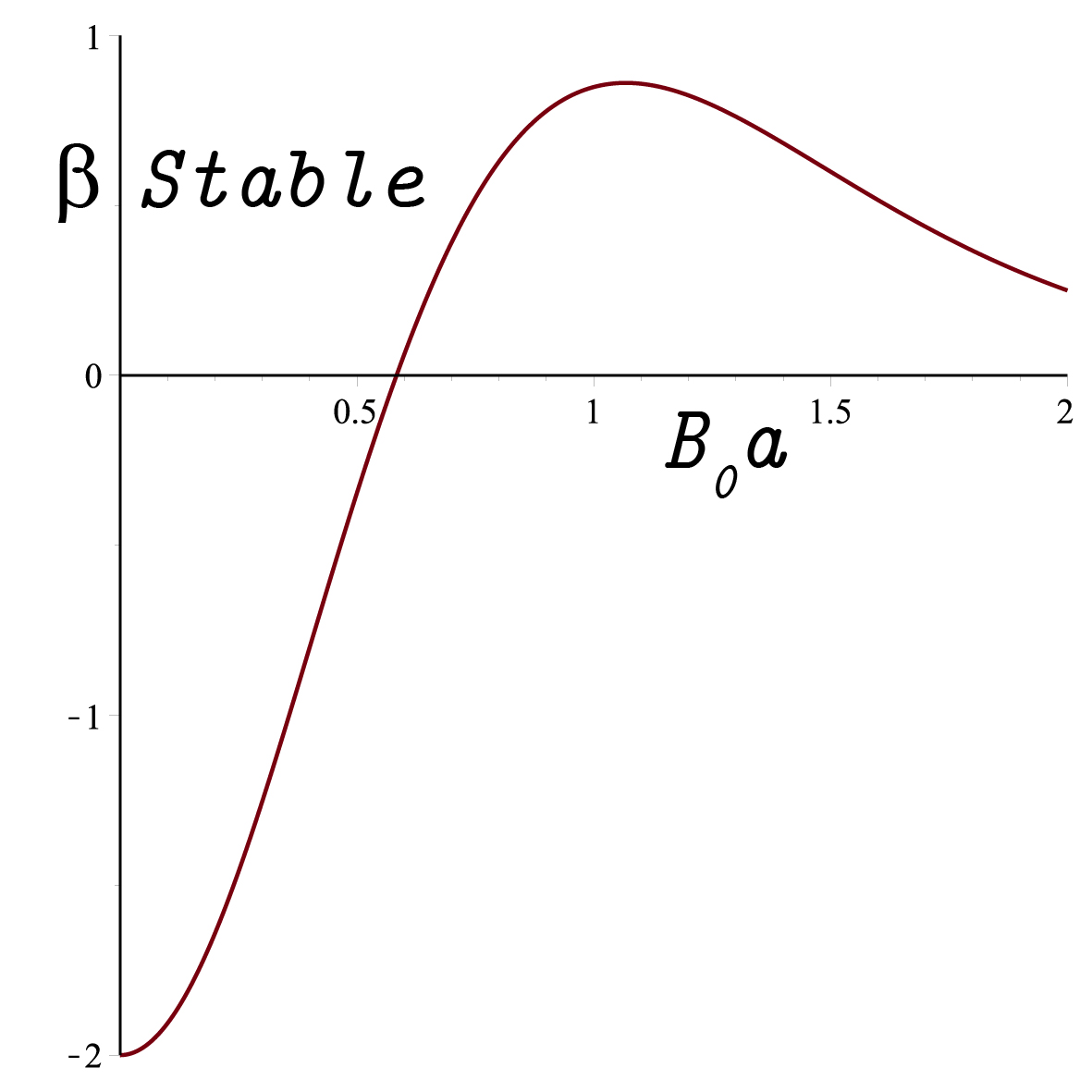}
\caption{Stability of TSW supported by CG in terms of $a_{0}B_{0}$ and $%
\protect\beta =\protect\beta _{1}=\protect\beta _{2}$. }
\end{figure}

Our second choice of the EoS is a CG. The form of $\Psi ^{\prime }$ and $%
\Phi ^{\prime }$ are given by 
\begin{equation}
\Psi ^{\prime }=\frac{\beta _{1}}{\sigma ^{2}}\text{ and }\Phi ^{\prime }=%
\frac{\beta _{2}}{\sigma ^{2}}
\end{equation}%
in which $\beta _{1}$ and $\beta _{2}$ are two new positive constants.
Furthermore, one finds%
\begin{equation}
\Psi \left( \sigma \right) =-\frac{\beta _{1}}{\sigma }+\Psi _{0}
\end{equation}%
and 
\begin{equation}
\Phi \left( \sigma \right) =-\frac{\beta _{2}}{\sigma }+\Phi _{0}
\end{equation}%
in which as before $\Psi _{0}$ and $\Phi _{0}$ are two integration
constants. Imposing the equilibrium conditions $\Psi \left( \sigma
_{0}\right) =P_{z0}$ and $\Phi \left( \sigma _{0}\right) =P_{\varphi 0}$\ we
find%
\begin{equation}
\Psi _{0}=P_{z0}+\frac{\beta _{1}}{\sigma _{0}}
\end{equation}%
and%
\begin{equation}
\Phi _{0}=P_{\varphi 0}+\frac{\beta _{2}}{\sigma _{0}}.
\end{equation}%
In Fig. 2 we plot the stability region of the TSW in terms of $\beta _{1}=$ $%
\beta _{2}=\beta $ and $B_{0}a.$ We note that setting $\beta _{1}=$ $\beta
_{2}=\beta $ makes $\Psi $ and $\Phi $ dependent as in the LG case i.e.,
(36) but in general they are independent.

\subsubsection{Generalized Chaplygin Gas (GCG)}

\begin{figure}[tbp]
\includegraphics[width=60mm,scale=0.7]{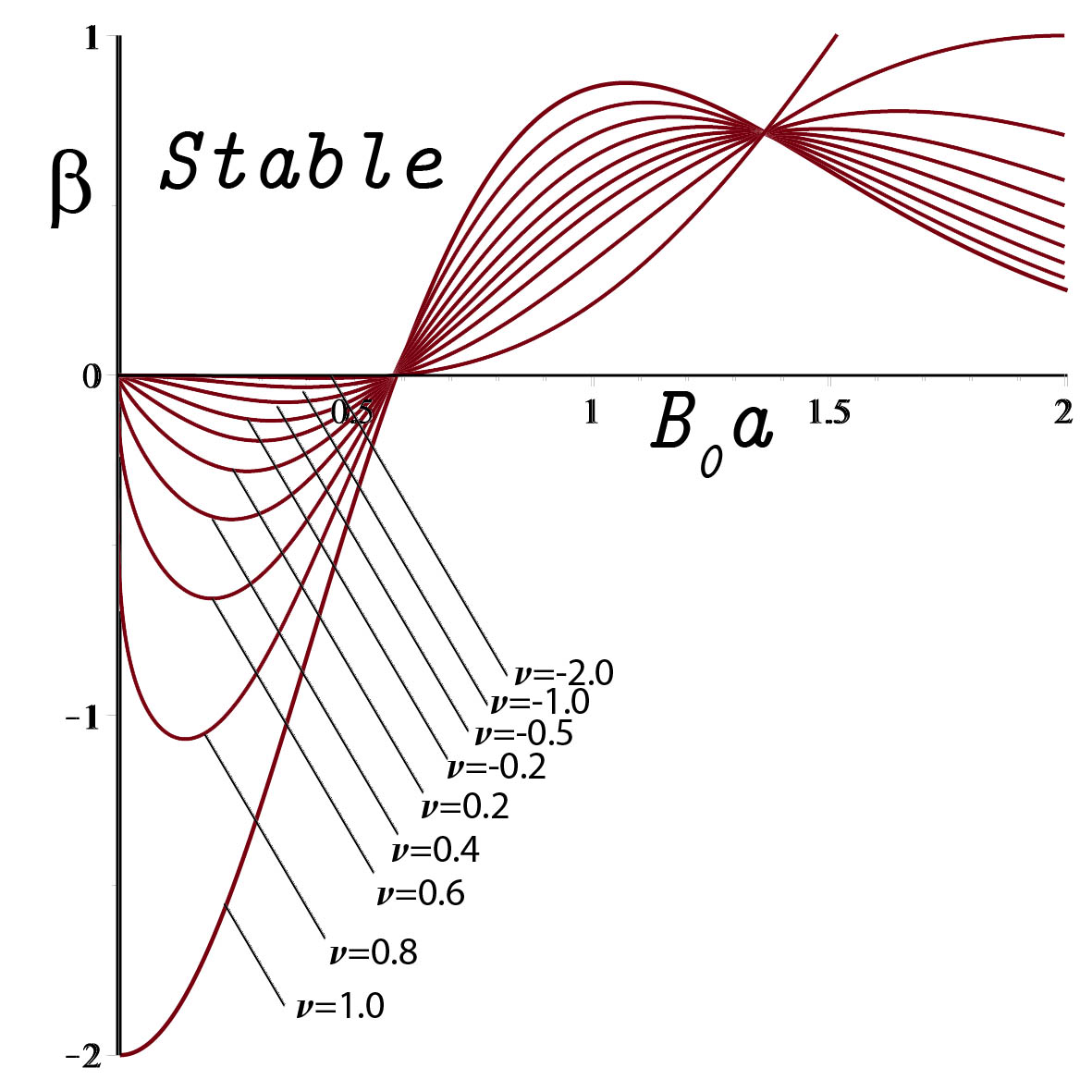}
\caption{Stability of TSW supported by GCG in terms of $a_{0}B_{0}$ and $%
\protect\beta =\protect\beta _{1}=\protect\beta _{2}$ with various value of $%
\protect\nu .$ The stable region is noted. }
\end{figure}

After CG in this part we consider a GCG EoS which is defined as 
\begin{equation}
\Psi ^{\prime }=\frac{\beta _{1}}{\sigma \left\vert \sigma \right\vert ^{\nu
}}\text{ and }\Phi ^{\prime }=\frac{\beta _{2}}{\sigma \left\vert \sigma
\right\vert ^{\nu }}
\end{equation}%
and consequently%
\begin{equation}
\Psi \left( \sigma \right) =-\frac{\beta _{1}}{\nu \left\vert \sigma
\right\vert ^{\nu }}+\Psi _{0}
\end{equation}%
and 
\begin{equation}
\Phi \left( \sigma \right) =-\frac{\beta _{2}}{\nu \left\vert \sigma
\right\vert ^{\nu }}+\Phi _{0}.
\end{equation}%
As before $\beta _{1}$ and $\beta _{2}$ are two new positive constants, $%
0<\nu \leq 1$ and $\Psi _{0}$ and $\Phi _{0}$ are integration constants. If
we set $\beta _{1}=$ $\beta _{2}=\beta $ again $\Psi $ and $\Phi $ are not
independent as Eq. (36). The equilibrium conditions imply 
\begin{equation}
\Psi _{0}=P_{z0}+\frac{\beta _{1}}{\nu \left\vert \sigma _{0}\right\vert
^{\nu }}
\end{equation}%
while%
\begin{equation}
\Phi _{0}=P_{\varphi 0}+\frac{\beta _{2}}{\nu \left\vert \sigma
_{0}\right\vert ^{\nu }}.
\end{equation}%
In Fig. 3 we show the effect of the additional freedom i.e., $\nu $ in the
stability of the corresponding TSW. We note that although in the standard
definition of the GCG one has to consider $0<\nu \leq 1$ in our figure we
also considered beyond this limit.

\subsubsection{Modified Generalized Chaplygin Gas (MGCG)}

\begin{figure}[tbp]
\includegraphics[width=60mm,scale=0.7]{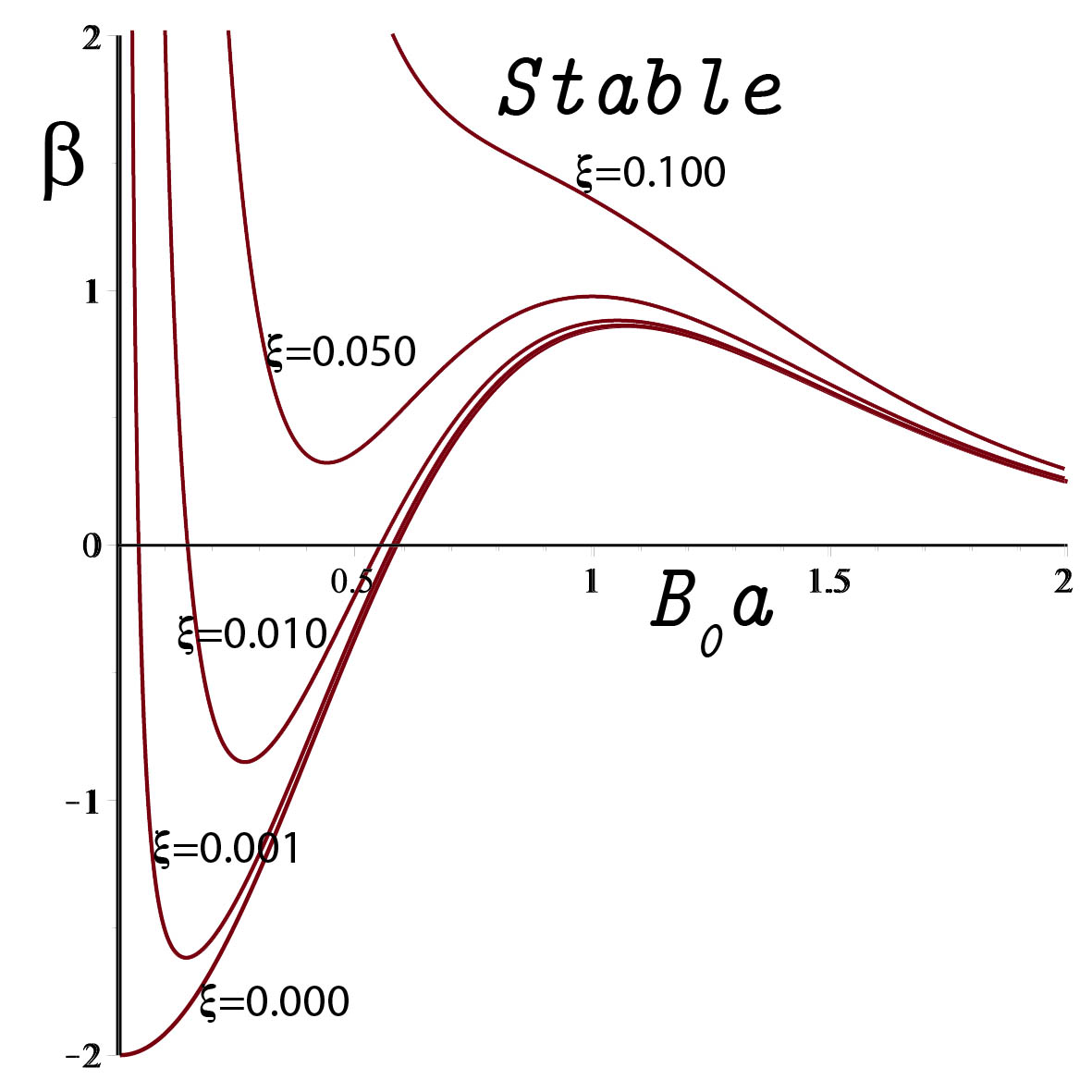}
\caption{Stability of TSW supported by MGCG in terms of $a_{0}B_{0}$ and $%
\protect\beta =\protect\beta _{1}=\protect\beta _{2}$. The different curves
are for different values of $\protect\xi =\protect\xi _{1}=\protect\xi _{2}$
and $\protect\nu $ is chosen to be $\protect\nu =1.$}
\end{figure}

Another step toward further generalization is to combine the LG and the GCG.
This is called MGCG and the form of the EoS may be written as%
\begin{equation}
\Psi ^{\prime }=\xi _{1}+\frac{\beta _{1}}{\sigma \left\vert \sigma
\right\vert ^{\nu }}\text{ and }\Phi ^{\prime }=\xi _{2}+\frac{\beta _{2}}{%
\sigma \left\vert \sigma \right\vert ^{\nu }}.
\end{equation}%
Herein, $\beta _{1}>0$, $\beta _{2}>0$, $\xi _{1}$ and $\xi _{2}$ are
constants and $0<\nu \leq 1.$ The form of $\Psi $ and $\Phi $ can be found as%
\begin{equation}
\Psi \left( \sigma \right) =\xi _{1}\sigma -\frac{\beta _{1}}{\nu \left\vert
\sigma \right\vert ^{\nu }}+\Psi _{0}
\end{equation}%
and%
\begin{equation}
\Phi \left( \sigma \right) =\xi _{2}\sigma -\frac{\beta _{2}}{\nu \left\vert
\sigma \right\vert ^{\nu }}+\Phi _{0}.
\end{equation}%
As before $\Psi _{0}$ and $\Phi _{0}$ are integration constants which can be
identified by imposing the similar equilibrium conditions i.e., $\Psi \left(
\sigma _{0}\right) =P_{z0}$ and $\Phi \left( \sigma _{0}\right) =P_{\varphi
0}.$ After that we find 
\begin{equation}
\Psi _{0}=P_{z0}+\frac{\beta _{1}}{\nu \left\vert \sigma _{0}\right\vert
^{\nu }}-\xi _{1}\sigma _{0}
\end{equation}%
and%
\begin{equation}
\Phi _{0}=P_{\varphi 0}+\frac{\beta _{2}}{\nu \left\vert \sigma
_{0}\right\vert ^{\nu }}-\xi _{2}\sigma _{0}.
\end{equation}%
In Fig. 4 we plot the stability region of the TSW supported by the MGCG with
additional arrangements as $\xi _{1}=\xi _{2}=\xi $ and $\beta _{1}=$ $\beta
_{2}=\beta .$ We again comment that these make $\Psi $ and $\Phi $ dependent
while in general they are independent. In Fig. 4 specifically we show the
effect of the additional freedom to the GCG, i.e., $\xi $ in a frame of $%
\beta $ and $B_{0}a.$

\subsubsection{Logarithmic Gas (LogG)}

\begin{figure}[tbp]
\includegraphics[width=60mm,scale=0.7]{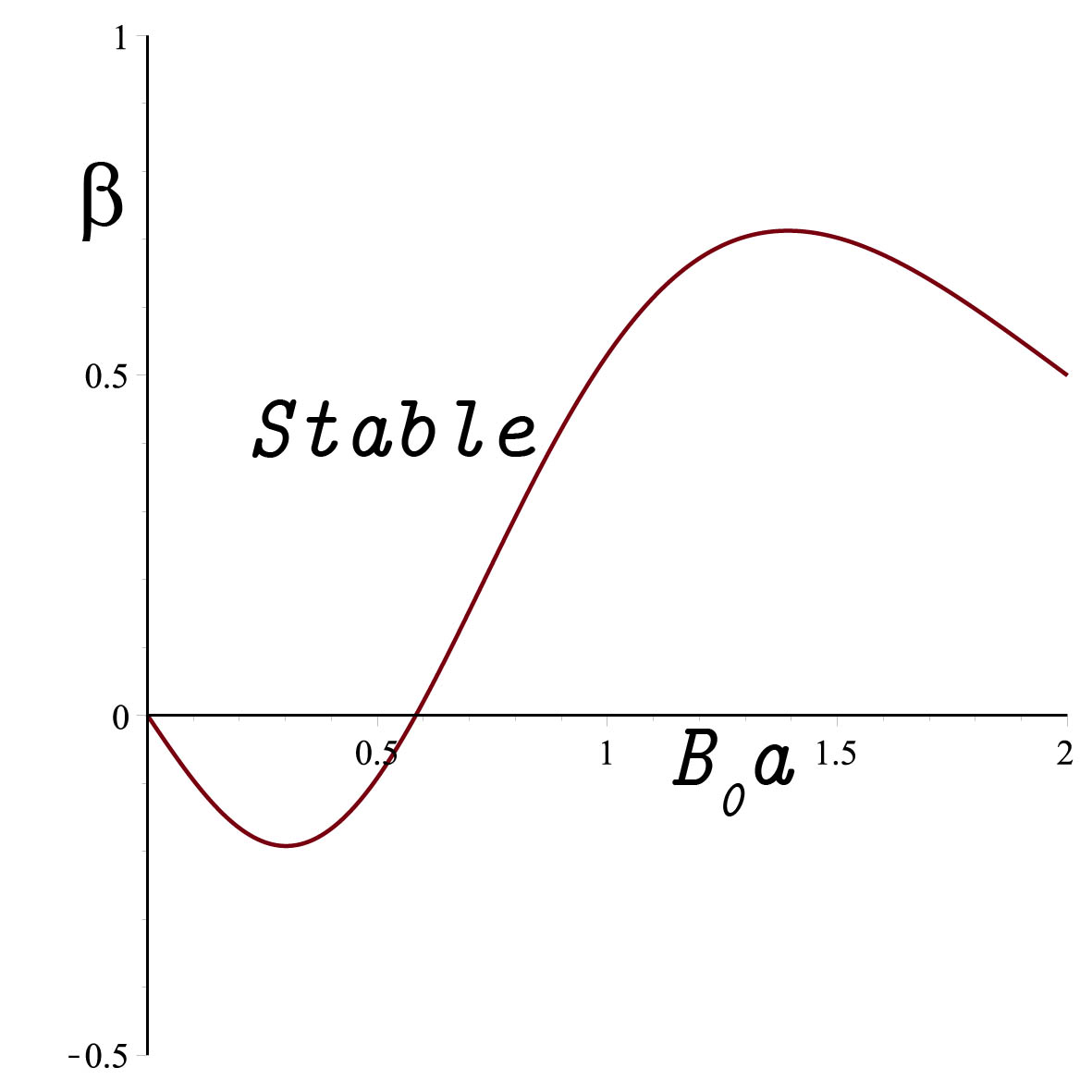}
\caption{Stability of TSW supported by LogG in terms of $a_{0}B_{0}$ and $%
\protect\beta =\protect\beta _{1}=\protect\beta _{2}$. We note that the
upper bound of $a_{0}B_{0}$ is chosen to be $2.$ }
\end{figure}

Finally we consider the LogG with 
\begin{equation}
\Psi ^{\prime }=-\frac{\beta _{1}}{\sigma }\text{ and }\Phi ^{\prime }=-%
\frac{\beta _{2}}{\sigma }
\end{equation}%
where $\beta _{1}>0$ and $\beta _{2}>0$ are two positive constants. The EoS
are given by%
\begin{equation}
\Psi =-\beta _{1}\ln \left\vert \frac{\sigma }{\sigma _{0}}\right\vert +\Psi
_{0}\text{ and }\Phi =-\beta _{1}\ln \left\vert \frac{\sigma }{\sigma _{0}}%
\right\vert +\Phi _{0}
\end{equation}%
in which the $\beta _{1}\ln \left\vert \sigma _{0}\right\vert +\Psi _{0}$
and $\beta _{2}\ln \left\vert \sigma _{0}\right\vert +\Phi _{0}$ are
integration constants. Imposing the equilibrium conditions one finds $\Psi
_{0}=P_{z0}$ and $\Phi _{0}=P_{\varphi 0}.$ In Fig. 5 we plot the stability
region in terms of $\beta _{1}=$ $\beta _{2}=\beta $ versus $B_{0}a.$

\section{Small velocity perturbation}

In the previous chapter we have considered a linear perturbation around the
equilibrium point of the throat. As we have considered above, the EoS of the
fluid on the thin shell after the perturbation had no relation with its
equilibrium state. However, by setting $\beta _{1}=\beta _{2}$ in our
analysis in previous chapter, implicitly we accepted that $\Psi -\Phi
=P_{z}-P_{\varphi }$ does not change in time, a restriction that is
physically acceptable.

In this chapter we consider the EoS of the TSW after the perturbation same
as its equilibrium point. This in fact means that the time evolution of the
throat is slow enough that any intermediate step between the initial point
and a certain final point can be considered as another equilibrium point (or
static). Quantitatively it means that $\frac{P_{z}}{\sigma }=-1$ (same as $%
\frac{P_{z0}}{\sigma _{0}}=-1$) and $\frac{P_{\varphi }}{\sigma }=-a\frac{%
U^{\prime }}{U}$ (same as$\frac{P_{\varphi 0}}{\sigma _{0}}=-a_{0}\frac{%
U_{0}^{\prime }}{U_{0}}$) and consequently, from (17), (18) and (19), we
find a single second order differential equation which may be written as%
\begin{equation}
2\ddot{a}+\frac{U^{\prime }}{U}\dot{a}^{2}=0.
\end{equation}%
This equation gives the exact motion of the throat after the perturbation.
(We note once more that the process of time evolution is considered with
small velocity). This equation can be integrated to obtain%
\begin{equation}
\dot{a}=\dot{a}_{0}\sqrt{\frac{U_{0}}{U}}.
\end{equation}%
A second integration with the exact form of $U,$ yields 
\begin{equation}
a\left( 1+\frac{B_{0}^{2}}{12}a^{2}\right) =a_{0}\left( 1+\frac{B_{0}^{2}}{12%
}a_{0}^{2}\right) +\dot{a}_{0}\sqrt{U_{0}}\left( \tau -\tau _{0}\right) .
\end{equation}%
The motion of the throat is under a negative force per unit mass which is
position and velocity dependent. As it is clear from the expression of $\dot{%
a},$ the magnitude of velocity is always positive and it never vanishes.
This means that the motion of the throat is not oscillatory but builds up in
the same direction after perturbation. Also from (56) we see that in proper
time if $\dot{a}_{0}>0,$ $a$ goes to infinity and when $\dot{a}_{0}<0,$ $a$
goes to zero. In both cases the particle-like motion does not return to its
initial position $a=a_{0}$. These mean that the TSW is not stable under
small velocity perturbations.

\section{TSW in Unified Bertotti-Robinson and Melvin spacetimes}

Recently two of us found a new solution to Einstein-Maxwell equations which
represents unified Bertotti-Robinson and Melvin spacetimes \cite{MH1} whose
line element is given by%
\begin{equation}
ds^{2}=-e^{2u}dt^{2}+e^{-2u}\left[ e^{2\kappa }\left( d\rho
^{2}+dz^{2}\right) +\rho ^{2}d\varphi ^{2}\right]
\end{equation}%
where 
\begin{equation}
e^{u}=F=\lambda _{0}\left[ \sqrt{\rho ^{2}+z^{2}}\cosh \left( \frac{B_{0}}{%
\lambda _{0}}\ln \rho \right) -z\sinh \left( \frac{B_{0}}{\lambda _{0}}\ln
\rho \right) \right] ,
\end{equation}%
and%
\begin{equation}
e^{\kappa }=\frac{F^{2}}{\left( \rho ^{2}+z^{2}\right) }\left[ \frac{\rho
^{1+\frac{B_{0}}{2\lambda _{0}}}}{z+\sqrt{\rho ^{2}+z^{2}}}\right] ^{\frac{%
2B_{0}}{\lambda _{0}}}.
\end{equation}%
Herein $\lambda _{0}$ and $B_{0}$ are two essential parameters of the
spacetime which are related to the magnetic field of the system and the
topology of the spacetime. The magnetic potential of the spacetime is given
by%
\begin{equation}
A_{\mu }=\Phi \left( \rho ,z\right) \delta _{\mu }^{\varphi }
\end{equation}%
in which 
\begin{equation}
\Phi _{\rho }\left( \rho ,z\right) =\rho e^{-2u}\psi _{z}
\end{equation}%
and%
\begin{equation}
\Phi _{z}\left( \rho ,z\right) =-\rho e^{-2u}\psi _{\rho }
\end{equation}%
with%
\begin{equation}
\psi =\lambda _{0}\left[ \sqrt{\rho ^{2}+z^{2}}\right] +B_{0}z.
\end{equation}%
The standard method of making TSW implies that $\mathcal{H}\left( \rho
\right) =\rho -a\left( \tau \right) =0$ is the timelike hypersurface where
the throat is located at and the line element on the shell reads%
\begin{equation}
ds^{2}=-d\tau ^{2}+e^{-2u\left( a,z\right) }\left[ e^{2\kappa \left(
a,z\right) }dz^{2}+a^{2}d\varphi ^{2}\right] .
\end{equation}%
The normal $4-$vector to the shell is found to be%
\begin{equation*}
n_{\gamma }^{\left( \pm \right) }=\pm \left( -\dot{a}e^{\kappa },e^{2\left(
\kappa -u\right) }\sqrt{\Delta },0,0\right) _{\Sigma },
\end{equation*}%
with $\Delta =\left( e^{2\left( u-\kappa \right) }+\dot{a}^{2}\right) $ and
the non-zero elements of the extrinsic curvature tensor become%
\begin{equation}
K_{\tau }^{\tau \left( \pm \right) }=\pm \left[ \frac{\ddot{a}+\left( \kappa
^{\prime }-u^{\prime }\right) \dot{a}^{2}}{\sqrt{\Delta }}+u^{\prime }\sqrt{%
\Delta }\right] ,
\end{equation}%
\begin{equation}
K_{z}^{z\left( \pm \right) }=\mp \left( u^{\prime }-\kappa ^{\prime }\right) 
\sqrt{\Delta },
\end{equation}%
and%
\begin{equation}
K_{\varphi }^{\varphi \left( \pm \right) }=\mp \left( u^{\prime }-\frac{1}{a}%
\right) \sqrt{\Delta }.
\end{equation}%
Upon the Israel junction conditions, one finds%
\begin{equation}
\sigma =2\sqrt{\Delta }\left( 2u^{\prime }-\kappa ^{\prime }-\frac{1}{a}%
\right) ,
\end{equation}%
\begin{equation}
P_{z}=2\left[ \frac{\ddot{a}+\left( \kappa ^{\prime }-u^{\prime }\right) 
\dot{a}^{2}}{\sqrt{\Delta }}+\frac{1}{a}\sqrt{\Delta }\right]
\end{equation}%
and%
\begin{equation}
P_{\varphi }=2\left[ \frac{\ddot{a}+\left( \kappa ^{\prime }-u^{\prime
}\right) \dot{a}^{2}}{\sqrt{\Delta }}+\kappa ^{\prime }\sqrt{\Delta }\right]
.
\end{equation}%
The results given above can be used to find the $\sigma _{0}$, $P_{z0}$ and $%
P_{\varphi 0}$ at the equilibrium radius $a=a_{0}$ i.e., 
\begin{equation}
\sigma _{0}=\left. 2e^{\left( u-\kappa \right) }\left( 2u^{\prime }-\kappa
^{\prime }-\frac{1}{a}\right) \right\vert _{a=a_{0}},
\end{equation}%
\begin{equation}
P_{z0}=\left. \frac{2}{a}e^{\left( u-\kappa \right) }\right\vert _{a=a_{0}}
\end{equation}%
and%
\begin{equation}
P_{\varphi 0}=\left. 2\kappa ^{\prime }e^{\left( u-\kappa \right)
}\right\vert _{a=a_{0}}.
\end{equation}%
Next, we use the exact form of $\kappa $ and $u$ to find the energy density
of the shell which can be written as%
\begin{equation}
\sigma _{0}=\frac{2a_{0}}{a_{0}^{2}+z^{2}}-\frac{\left( \epsilon +1\right)
^{2}}{a_{0}}+\frac{2\epsilon a_{0}}{\sqrt{a_{0}^{2}+z^{2}}\left( z+\sqrt{%
a_{0}^{2}+z^{2}}\right) }
\end{equation}%
in which $\epsilon =\frac{B_{0}}{\lambda _{0}}.$ To analyze the sign of $%
\sigma _{0}$ we introduce $\zeta =\frac{z}{a_{0}}$ and rewrite the latter
equation as%
\begin{equation}
a_{0}\sigma =-\left( 1+\epsilon \right) ^{2}+\frac{2\left( \zeta +\left(
1+\epsilon \right) \sqrt{1+\zeta ^{2}}\right) }{\left( 1+\zeta ^{2}\right)
\left( \zeta +\sqrt{1+\zeta ^{2}}\right) }.
\end{equation}%
One of the interesting case is when we set $\epsilon =-1$ which yields%
\begin{equation}
a_{0}\sigma _{0}=\frac{2\zeta }{\left( 1+\zeta ^{2}\right) \left( \zeta +%
\sqrt{1+\zeta ^{2}}\right) }.
\end{equation}%
This is positive for $\zeta >0$ ($z>0$), negative for $\zeta <0$ ($z<0$) and
zero for $\zeta =0$ ($z=0$)$.$ Another interesting case is when we set $%
\epsilon =0$ which is the BR limit of the general solution (57-59). In this
setting we find%
\begin{equation}
a_{0}\sigma _{0}=\frac{2}{1+\zeta ^{2}}-1
\end{equation}%
which is positive for $\left\vert \zeta \right\vert <1.$ In Fig. 6 we plot
the region on which $a_{0}\sigma _{0}\geq 0$ in terms of $\epsilon $ and $%
\zeta .$ To find the total energy of the shell we use%
\begin{equation}
\Omega =\int_{0}^{2\pi }\int_{-\infty }^{+\infty }\int_{0}^{\infty }\sigma
_{0}\delta \left( \rho -a_{0}\right) \sqrt{-g}d\rho dzd\varphi
\end{equation}%
which after some manipulation becomes%
\begin{equation}
\Omega =2\pi \int_{-\infty }^{+\infty }\sigma _{0}a_{0}e^{2\left( \kappa
_{0}-u_{0}\right) }dz.
\end{equation}%
in which $\kappa _{0}=\left. \kappa \right\vert _{a=a_{0}}$ and $%
u_{0}=\left. u\right\vert _{a=a_{0}}.$%
\begin{figure}[tbp]
\includegraphics[width=60mm,scale=0.7]{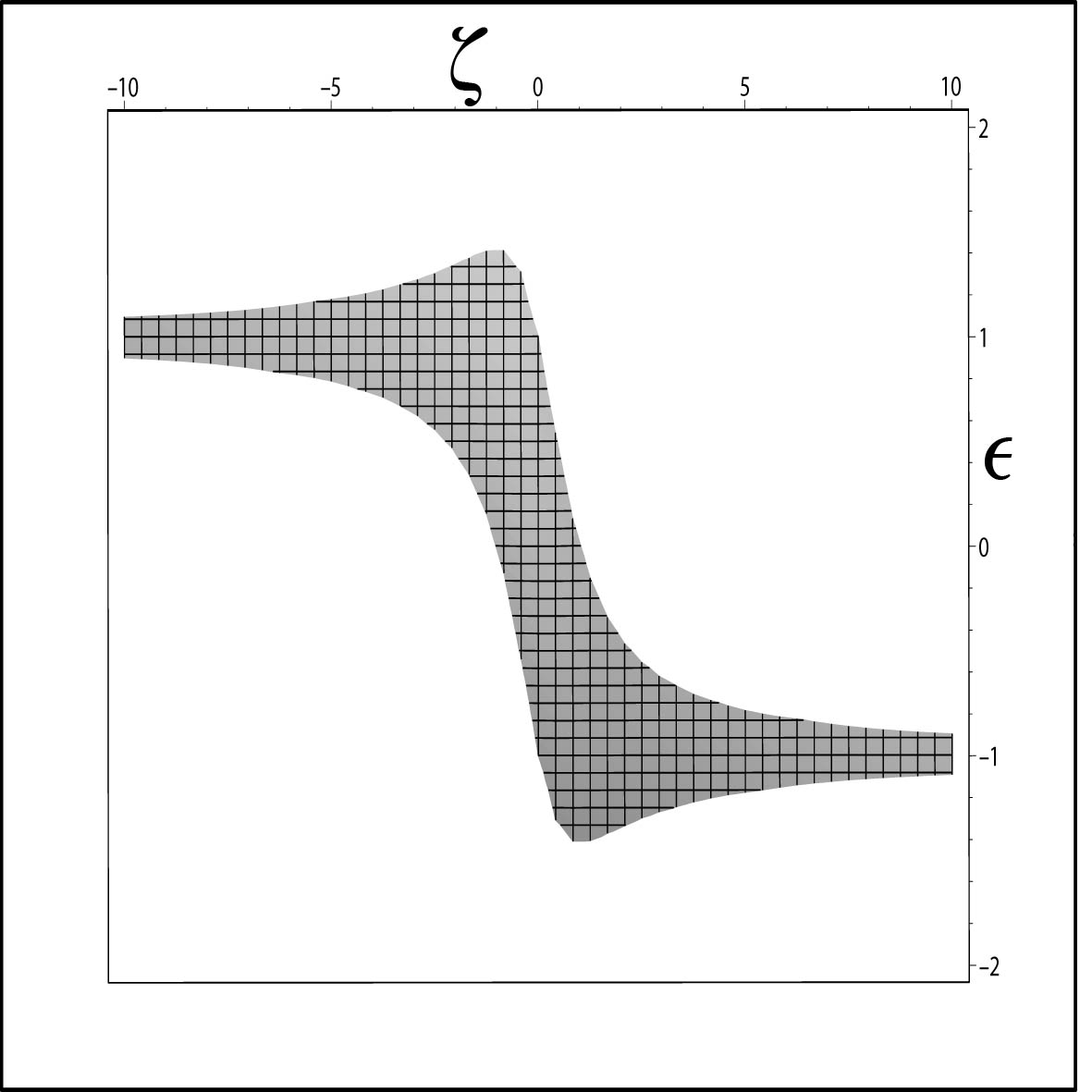}
\caption{$a_{0}\protect\sigma _{0}$ versus $\protect\epsilon $ and $\protect%
\zeta .$ The shaded region in the region on which $a_{0}\protect\sigma _{0}$
is positive. }
\end{figure}
Upon some further manipulation we arrive at%
\begin{multline}
\frac{\Omega }{2\pi \lambda _{0}^{2}a_{0}^{2\epsilon ^{2}-1}}= \\
\int_{-\infty }^{\infty }\left[ \frac{2\left( \zeta +\left( 1+\epsilon
\right) \sqrt{1+\zeta ^{2}}\right) }{\left( 1+\zeta ^{2}\right) ^{3}\left(
\zeta +\sqrt{1+\zeta ^{2}}\right) }-\frac{\left( 1+\epsilon \right) ^{2}}{%
\left( 1+\zeta ^{2}\right) ^{2}}\right] \\
\frac{\left( \sqrt{1+\zeta ^{2}}\cosh \left( \epsilon \ln a_{0}\right)
-\zeta \sinh \left( \epsilon \ln a_{0}\right) \right) ^{2}}{\left( \zeta +%
\sqrt{1+\zeta ^{2}}\right) ^{4\epsilon }}d\zeta .
\end{multline}%
Although this integral can not be evaluated explicitly for arbitrary $%
\epsilon $ at least for $\epsilon =0$ it gives%
\begin{equation}
\Omega =\lim_{R\rightarrow \infty }\frac{4\pi \lambda _{0}^{2}R}{a_{0}\left(
1+R^{2}\right) }
\end{equation}%
which is positive. Obviously this limit (i.e. $\epsilon =0$) corresponds to
the Bertotti-Robinson limit of the general solution in which for $R<\infty $
construction of a TSW with a positive total energy becomes possible.

\section{Conclusion}

A large class of stable TSW solutions is found by employing the magnetic
Melvin universe through the cut-and-paste technique. The Melvin spacetime is
a typical cylindrically symmetric, regular solution of the Einstein-Maxwell
equations. Herein the throat radius of the TSW is confined by a strong
magnetic field, for this reason we phrase them as \textit{microscopic
wormholes}. Being regular its construction can be achieved by a finite
energy. It has recently been suggested that the mysterious EPR particles may
be connected through a wormhole \cite{Sus}. From this point of view the
magnetic Melvin wormhole may be instrumental to test such a claim. We have
applied radial, linear perturbation to the throat radius of the TSW in
search for stability regions. In such perturbations we observed that the
initial radial speed must be chosen zero in order to attain a stable TSW.
Different perturbations may cause collapse of the wormhole. As the material
on the throat we have adopted various equations of states, ranging from an
ordinary linear / logarithmic gas to a Chaplygin gas. The repulsive support
derived from such sources gives life to the TSW against the gravitational
collapse. Besides pure Melvin case we have also considered TSW in the
magnetic universe of unified Melvin and Bertotti-Robinson spacetimes. The
pure Bertotti-Robinson TSW has positive total energy for each finite axial
length ($R<\infty $). The energy becomes zero when the cut-off length $%
R\rightarrow \infty .$


\begin{thebibliography}{99}
\bibitem{Melvin} M.\thinspace A. Melvin, Phys. Lett. \textbf{8}, 65 (1964);

W.\thinspace B. Bonnor, Proc. Phys. Soc. London Sect. A \textbf{67}, 225
(1954);

D. Garfinkle and E.\thinspace N. Glass, Classical Quantum Gravity \textbf{28}%
, 215012 (2011).

\bibitem{TSWS} M. Visser, Phys. Rev. D \textbf{39}, 3182 (1989);

M. Visser, Nucl. Phys. \textbf{B} 328, 203 (1989);

P. R. Brady, J. Louko and E. Poisson, Phys. Rev. D \textbf{44}, 1891 (1991);

E. Poisson and M. Visser, Phys. Rev. D \textbf{52}, 7318 (1995);

M. Ishak and K. Lake, Phys. Rev. D \textbf{65}, 044011 (2002);

C. Simeone, Int. Jou. of Mod. Phys. D \textbf{21}, 1250015 (2012);

F. S. N. Lobo, Phys. Rev. D \textbf{71}, 124022 (2005);

E. F. Eiroa and C. Simeone, Phys. Rev. D \textbf{71}, 127501 (2005);

E. F. Eiroa, Phys. Rev. D \textbf{78}, 024018 (2008);

F. S. N. Lobo and P. Crawford, Class. Quantum Grav. \textbf{22,} 4869 (2005);

S. H. Mazharimousavi, M. Halilsoy and Z. Amirabi, Phys. Lett. A \textbf{375}%
, 3649 (2011);

M. Sharif and M. Azam, Eur. Phys. J. C \textbf{73}, 2407 (2013);

M. Sharif and M. Azam, Eur. Phys. J. C \textbf{73}, 2554 (2013);

S. H. Mazharimousavi and M. Halilsoy, Eur. Phys. J. C \textbf{73}, 2527
(2013).

\bibitem{TSWC} E. F. Eiroa and C. Simeone, Phys. Rev. D \textbf{70}, 044008
(2004);

M. Sharif and M. Azam, JCAP \textbf{04}, 023 (2013);

E. Rub\'{\i}n de Celis, O. P. Santillan and C. Simeone, Phys. Rev. D \textbf{%
86}, 124009 (2012);

C. Bejarano, E. F. Eiroa and C. Simeone, Phys. Rev. D \textbf{75}, 027501
(2007);

K. A. Bronnikov, V. G. Krechet and J. P. S. Lemos, Phys. Rev. D \textbf{87},
084060 (2013);

M. G. Richarte, Phys. Rev. D \textbf{87}, 067503 (2013);

Z. Amirabi, M. Halilsoy and S. H. Mazharimousavi, Phys. Rev. D \textbf{88},
124023 (2013).

\bibitem{EPR} A. Einstein, B Podolsky and N. Rosen, Phys. Rev. \textbf{47},
777 (1935).

\bibitem{MH1} S. H. Mazharimousavi and M. Halilsoy, Phys. Rev. D \textbf{88}%
, 064021 (2013).

\bibitem{BR} K. A. Bronnikov and J. P. S. Lemos, Phys. Rev. D \textbf{79},
104019 (2009).

\bibitem{ES1} E. F. Eiroa and C. Simeone, Phys. Rev. D \textbf{81}, 084022
(2010).

\bibitem{ES2} E. F. Eiroa and C. Simeone, Phys. Rev. D \textbf{82}, 084039
(2010).

\bibitem{R} M. G. Richarte, Phys. Rev. D \textbf{88}, 027507 (2013).

\bibitem{Visser} M. Visser, Phys. Rev. D \textbf{39,} 3182 (1989);

M. Visser, Nucl. Phys. B \textbf{328}, 203 (1989).

\bibitem{Israel} W. Israel, Nuovo Cimento \textbf{44B}, 1 (1966);

V. de la Cruzand W. Israel, Nuovo Cimento \textbf{51A}, 774 (1967);

J. E. Chase, Nuovo Cimento \textbf{67B}, 136. (1970);

S. K. Blau, E. I. Guendelman, and A. H. Guth, Phys. Rev. D \textbf{35}, 1747
(1987);

R. Balbinot and E. Poisson, Phys. Rev. D \textbf{41}, 395 (1990).

\bibitem{MH2} S. Habib Mazharimousavi, M. Halilsoy and Z. Amirabi, Phys.
Rev. D (2014) in press, arXiv:1403.2861.

\bibitem{Sus} J. Maldacena and L. Susskind, arXiv:1306.0533 "\textit{Cool
horizons for entangled black holes}".
\end{thebibliography}
\end{document}